%% file: main.tex
\documentclass[submission,copyright,creativecommons]{eptcs}
\providecommand{\event}{F-IDE 2018} % Name of the event you are submitting to
\usepackage{underscore}           % Only needed if you use pdflatex.

\usepackage[T1]{fontenc}

\usepackage{amssymb}
\usepackage{graphicx}
\usepackage{verbatim}
\usepackage{xcolor}
\usepackage{varioref}
\usepackage{paralist}
\usepackage{amssymb}

\newcommand{\ifnotdraft}[1]{\ifx\@draft\@undefined #1 \fi}
\newcommand{\ifnotdraftelse}[2]{\ifx\@draft\@undefined #1 \else #2 \fi}
\newcommand{\ifdraft}[1]{\ifx\@draft\@undefined  \else #1 \fi}
\newcommand{\ifprint}[2]{\ifx\@print\@undefined #2 \else #1 \fi}
\newcommand{\iffinal}[2]{\ifx\@final\@undefined #2 \else #1 \fi}

\input{macros/colorscheme}
\input{macros/macros}
\input{macros/lststyles}

\usepackage[normalem]{ulem}

\def\lstsmallmath{\leavevmode\ifmmode \scriptstyle \else  \fi}
\def\lstsmallmathend{\leavevmode\ifmmode  \else  \fi}

\usepackage{xspace}
\newcommand\eg{e.g.,\xspace}
\newcommand\ie{i.e.,\xspace}

\usepackage{url}
\urldef{\mailsa}\path|{a.knueppel, t.thuem, carsten.burmeister, i.schaefer}@tu-bs.de|

\usepackage{adjustbox}

\title{Experience Report on Formally\\ Verifying Parts of OpenJDK's API with KeY}
\author{Alexander Kn\"uppel,  Thomas Th\"um, Carsten Pardylla, and Ina Schaefer
\institute{TU Braunschweig, Germany}
\email{\{a.knueppel, t.thuem, carsten.burmeister, i.schaefer\}@tu-bs.de}
}

\def\titlerunning{Experience Report on Formally Verifying Parts of OpenJDK's API with KeY}
\def\authorrunning{A. Kn\"uppel, T. Th\"um, C. Pardylla, and I. Schaefer}

\begin{document}
\maketitle

\begin{abstract}
Deductive verification of software has not yet found its way into industry, as complexity and scalability issues require highly specialized experts. The long-term perspective is, however, to develop verification tools aiding industrial software developers to find bugs or bottlenecks in software systems faster and more easily. The KeY project constitutes a framework for specifying and verifying software systems, aiming at making formal verification tools applicable for mainstream software development. 
To help the developers of KeY, its users, and the deductive verification community, we summarize our experiences with KeY 2.6.1 in specifying and verifying real-world Java code from a users perspective. To this end, we concentrate on parts of the Collections-API of OpenJDK~6, where an informal specification exists. While we describe how we bridged informal and formal specification, we also exhibit accompanied challenges that we encountered. 
%As a foundation, we used exhibit
%Since KeY does not support a higher version than Java 1.2, we either modified classes with unsupported language constructs or sliced such classes completely. 
%Our experiences are that (a) in principle, deductive verification is feasible, but requires high expertise, (b) developing formal specifications for existing code bases is notoriously hard, and (c) KeY is unsound due to Java's misconception of the maximal array length. 
Our experiences are that (a) in principle, deductive verification for API-like code bases is feasible, but requires high expertise, (b) developing formal specifications for existing code bases is still notoriously hard, and (c) the under-specification of certain language constructs in Java is challenging for tool builders. Our initial effort in specifying parts of OpenJDK~6 constitutes a stepping stone towards a case study for future research.

%that specification is currently the bottleneck in deductive verification, and that the influence of particular configuration options is difficult to understand for non-experts.
%Our insights are not depending on KeY in particular, but can be generalized to most verification tools currently researched or in development.

%\keywords{deductive verification, design by contract, formal methods, theorem proving, KeY}
\end{abstract}

\lstset{style=java,style=featurehouse,style=jml,tabsize=2,escapechar=|}
\labelformat{lstlisting}{Listing~#1}

\renewcommand{\thelstlisting}{\arabic{lstlisting}}

\input{content/introduction}
\input{content/contracts}

\input{content/specifying-collection}
\input{content/experiences}
\input{content/related}
\input{content/conclusion}

%% Acknowledgements
\subsubsection*{Acknowledgments.}
We gratefully acknowledge Richard Bubel, Reiner H\"ahnle, Dominik Steinh\"ofel, and Stefan Kr\"uger for fruitful discussions and valuable feedback throughout this work. Work on this paper was partially supported by the DFG Research Unit \emph{"Controlling Concurrent Change"}, funding number FOR 1800. 

%\bibliographystyle{abbrv}
%\small
%\nocite{*}
\bibliographystyle{eptcs}
\bibliography{MYabrv,literature}

\end{document}

%% file: macros/colorscheme.tex
\definecolor{blue1}{RGB}{0,105,180}%95
\definecolor{blue2}{RGB}{40,87,121}
\definecolor{blue3}{RGB}{0,57,97}%51
\definecolor{blue4}{RGB}{76,166,230}%157
\definecolor{blue5}{RGB}{136,191,230}%186

\definecolor{orange1}{RGB}{255,144,0}%133
\definecolor{orange2}{RGB}{171,121,56}
\definecolor{orange3}{RGB}{137,78,0}%72
\definecolor{orange4}{RGB}{255,181,84}
\definecolor{orange5}{RGB}{255,210,151}%205

\definecolor{green1}{RGB}{11,215,0}%75
\definecolor{green2}{RGB}{52,144,48}
\definecolor{green3}{RGB}{6,116,0}%41
\definecolor{green4}{RGB}{88,241,80}
\definecolor{green5}{RGB}{148,241,143}%177

\definecolor{red1}{RGB}{253,0,6}%86
\definecolor{red2}{RGB}{170,56,59}
\definecolor{red3}{RGB}{136,0,3}%46
\definecolor{red4}{RGB}{254,84,88}
\definecolor{red5}{RGB}{254,151,154}%186

\definecolor{gray1}{RGB}{97,97,97}
\definecolor{gray2}{RGB}{94,94,94}
\definecolor{gray3}{RGB}{53,53,53}
\definecolor{gray4}{RGB}{152,152,152}
\definecolor{gray5}{RGB}{189,189,189}

\definecolor{background}{named}{white}
\definecolor{bgborder}{named}{black}
\definecolor{comment}{named}{red3}

\definecolor{blue}{named}{blue1}
\definecolor{green}{named}{green1}
\definecolor{red}{named}{red1}
\definecolor{orange}{named}{orange1}

\definecolor{coqcomment}{named}{comment}
\definecolor{coqred}{named}{black}
\definecolor{coqblue}{named}{black}
\definecolor{acolor1}{named}{red1}
\definecolor{acolor2}{named}{blue1}
\definecolor{gacolor1}{gray}{0.34}%86
\definecolor{gacolor2}{gray}{0.37}%95
\definecolor{highlight}{gray}{0.9}

\definecolor{pdflinkcolor}{named}{blue3}
\definecolor{pdfcitecolor}{named}{green3}

\definecolor{grayed}{gray}{0.66}

%% file: macros/macros.tex
\usepackage{amssymb}
\usepackage{booktabs} % toprule, midrule, bottomrule

\usepackage{array} % tabular m{1cm}
\usepackage{ragged2e} % own column types P{1cm}
\newcolumntype{P}[1]{>{\RaggedRight\hspace{0pt}}p{#1}}
\newcolumntype{M}[1]{>{\RaggedRight\hspace{0pt}}m{#1}}

\newcommand{\code}[1]{\mbox{\sf#1}}

\newcommand{\KeY}{Ke\kern-0.1emY}

%% file: macros/lststyles.tex
\usepackage{listings} % source code listings

\newcommand{\modsep}{\hrule\vspace{1mm}\hrule}
\ifprint{
	\newcommand{\jmlkeyword}[1]{\color{darkgray}\textbf{#1}}
}{
	\newcommand{\jmlkeyword}[1]{\color{green3}\textbf{#1}}
}
\newcommand{\invariant}{\jmlkeyword{invariant}}
\newcommand{\public}{\jmlkeyword{public}}

\newcommand{\normalbehavior}{\jmlkeyword{normal\_behavior}}
\newcommand{\exceptionalbehavior}{\jmlkeyword{exceptional\_behavior}}

\newcommand{\requires}{\jmlkeyword{requires}}
\newcommand{\assignable}{\jmlkeyword{assignable}}
\newcommand{\ensures}{\jmlkeyword{ensures}}
\newcommand{\also}{\jmlkeyword{also}}
\newcommand{\signals}{\jmlkeyword{signals}}
\newcommand{\signalsonly}{\jmlkeyword{signals\_only}}
\newcommand{\modeljml}{\jmlkeyword{model}}

\newcommand{\nullablejml}{\jmlkeyword{nullable}}

\ifprint{
	\newcommand{\jmlkeywordb}[1]{\color{darkgray}\textbackslash\textbf{#1}}
}{
	\newcommand{\jmlkeywordb}[1]{\color{green3}\textbackslash\textbf{#1}}
}

\newcommand{\foralljml}{\jmlkeywordb{forall}}
\newcommand{\old}{\jmlkeywordb{old}}
\newcommand{\typeof}{\jmlkeywordb{typeof}}
\newcommand{\result}{\jmlkeywordb{result}}

\newcommand{\fresh}{\jmlkeywordb{fresh}}

\lstdefinestyle{java}{
	language=Java,
	tabsize=4,
	breaklines=false,
	basicstyle=\sf\small\selectfont,
	commentstyle=\fontshape{it}\color{darkgray}\selectfont,
	keywordstyle=\fontseries{b}\selectfont,%\color{orange3}
	stringstyle=\selectfont,
	columns=fullflexible, 
	numbers=none,%left, right, none
	numberstyle=\footnotesize,
	captionpos=b,
	frame=single,%trblTRBL
	framesep=3pt,
	xleftmargin=4pt,
	xrightmargin=4pt,
	rulecolor=\color{bgborder},
	escapechar=|,
	firstnumber=auto,
	showstringspaces=false,
}

\lstdefinestyle{xml}{
	style=java,
	tabsize=2,
	language=xml,
}

\lstdefinestyle{featureidexml}{
	style=xml,
	morekeywords={featureModel, struct, feature, name, abstract, mandatory, and, alt, or, constraints, rule, imp, var, comments},
}

\lstdefinestyle{jak}{
	style=java,
	morekeywords={layer, refines, Super(), Super},
}

\lstdefinestyle{featurehouse}{
	style=java,
	morekeywords={original, refines},
}

\lstdefinestyle{deltaj}{
	style=java,
	morekeywords={original, core, delta, when, after, before, modifies, adds, removes},
}

\lstdefinestyle{aspectj}{
	style=java,
	morekeywords={aspect, after, call},
}

\lstdefinestyle{ifdef}{
  morecomment=[l]{\#},
}

\lstdefinestyle{interface}{
	morekeywords={requires, provides},
}

\ifprint{
	\lstdefinestyle{jml}{
		commentstyle=\color{darkgray}\selectfont,%\fontshape{it}
	}
}{
	\lstdefinestyle{jml}{
		commentstyle=\color{green3}\selectfont,%\fontshape{it}
	}
}

\lstdefinestyle{small}{
	basicstyle=\sf\small\selectfont
}
\lstdefinestyle{footnotesize}{
	basicstyle=\sf\footnotesize\selectfont
}
\lstdefinestyle{scriptsize}{
	basicstyle=\sf\scriptsize\selectfont
}
\lstdefinestyle{tiny}{
	basicstyle=\sf\tiny\selectfont
}

\lstdefinelanguage{Coq}{
  morekeywords={Variable,Inductive,CoInductive,Fixpoint,CoFixpoint,%
      Definition,Lemma,Theorem,Axiom,Local,Save,Grammar,Syntax,Intro,%
      Trivial,Qed,Intros,Symmetry,Simpl,Rewrite,Apply,Elim,Assumption,%
      Left,Cut,Auto,Unfold,Exact,Right,Hypothesis,Pattern,Destruct,%
      Constructor,Defined,Fix,Record,Proof,Induction,Hints,Exists,let,in,%
      Parameter,Split,Red,Reflexivity,Transitivity,if,then,else,Opaque,%
      Transparent,Inversion,Absurd,Generalize,Mutual,Cases,of,end,Analyze,%
      AutoRewrite,Functional,Scheme,params,Refine,using,Discriminate,Try,%
      Require,Load,Import,Scope,Set,Open,Section,End,match,with,Ltac,%
			exists,forall
	},% 
  sensitive, %
  morecomment=[n]{(*}{*)},%
  morestring=[d]",%
  literate=
   {=>}{{$\Rightarrow$}}1
	 {>->}{{$\rightarrowtail$}}2
	 {<->}{{$\leftrightarrow$}}2
	 {->}{{$\to$}}1
    {\/\\}{{$\wedge$}}1
    {|-}{{$\vdash$}}1
    {\\\/}{{$\vee$}}1
    {~}{{$\sim$}}1
}[keywords,comments,strings]%

\lstdefinestyle{coqblack}{
	language=Coq,
	tabsize=4,
	breaklines=false,
	basicstyle=\sf\small\selectfont,
	commentstyle=\fontshape{it}\selectfont,
	keywordstyle=\fontseries{b}\selectfont,%\color{orange3}
	stringstyle=\selectfont,
	columns=fullflexible, 
	numbers=none,
	numberstyle=\footnotesize,
	captionpos=b,
	frame=single,%trblTRBL
	xleftmargin=4pt,
	xrightmargin=4pt,
	rulecolor=\color{bgborder},
	morekeywords={Module, Fact, Export, Tactic, Notation, Prop, Combined, Variables},
  emph={Variable, Hypothesis, Module, Fact, Lemma, Theorem, Ltac, Inductive, Fixpoint, Definition, Parameter, with, admit, Variables},
	emphstyle={\fontseries{b}\selectfont},
  escapeinside={ß}{ß} % escapeinside={$}{$}
}

\ifprint{
	\lstdefinestyle{coq}{
		style=coqblack
	}
}{
	\lstdefinestyle{coq}{
		style=coqblack,
		commentstyle=\fontshape{it}\color{darkgray}\selectfont,
		keywordstyle=\fontseries{b}\color{blue3}\selectfont,%\color{orange3}
		emphstyle={\color{red3}\fontseries{b}\selectfont},
	}
}

\lstdefinestyle{slides}{
	backgroundcolor=\color{white}
}

\usepackage{fancyvrb}

\makeatletter
\newenvironment{savelst}{\VerbatimEnvironment\begin{VerbatimOut}{vsave.tmp}}
	{\end{VerbatimOut}}

\newcommand{\savedlst}{\lstinputlisting{vsave.tmp}}

\newcommand{\savedlstbox}{\settowidth\@tempdima{%
		{\renewcommand{\pause}{}  \savedlst}}
	\minipage{\@tempdima}\savedlst\endminipage}

\makeatother

\lstdefinestyle{customc}{
	language=C,
	frame=single,
	framerule=0.9pt,
	framesep=9pt,
	basicstyle=\ttfamily\bfseries,
	commentstyle=\color{green!40!black},
	identifierstyle=\color{red},
	directivestyle=\color{gray},
	stringstyle=\color{orange},
	escapechar=|,
	tabsize=4,
}

\lstdefinestyle{customjava}{
	language=Java,
	frame=single,
	framerule=0.9pt,
	framesep=9pt,
	basicstyle=\ttfamily\bfseries,
	keywordstyle=\color{blue},
	commentstyle=\color{green!40!black},
	identifierstyle=\color{black},
	stringstyle=\color{orange},
	escapechar=|,
	tabsize=4,
}

%% file: content/introduction.tex
\section{Introduction}
\label{sec:introduction}

% #1 general introduction to deductive verification on source code level with key

% #2 highly dependet on user interaction -> only suitable for experts

% #3 In particular, deductive verification of legacy systems is hard, because they are not designed for verification in the first place

% #4 But using verification tools is generally hard

% #5 Contribution of this paper is two-fold:
%     - Scenario-based discussion on 

In particular for safety-critical systems, the main goal of formal methods is to increase confidence in the correctness of a software system by providing means to mechanically reason about it~\cite{BS:SEJ93,CW:CSUR96,R97,SD88}. Driven by research, formal methods have experienced significant advancements over the last decades. Besides lightweight methods like testing and code reviews, formal verification techniques, such as deductive verification~\cite{BFL+11,BHS07,BCC+05,HLL+12,schumann2001automated} or model checking~\cite{CGP99,RRDH06}, can be used to ensure that a program \emph{behaves correctly} by proving that it adheres to a formal specification. Recent successes even show that verifying large-scale software systems written in real-world programming languages is not only theoretically, but indeed practically feasible~\cite{baumann2012lessons}, and that bugs in real-world software used by millions of devices can be identified~\cite{de2015openjdk}.  

A downside is that formal verification has not yet found its way into industrial software development.
One reason is the substantial specification effort, which is particularly uneconomical for legacy software systems~\cite{beckert2016deductive}. In particular, there exist serious doubts about the cost-effectiveness (\ie return of investement) of formal methods~\cite{knight1997formal}. Another reason is that developers regard most formal methods as difficult to
understand and apply, as full automation is often impossible due to the undecidability of the halting problem~\cite{rogers1967theory}. During the verification phase, there is often a demand for user interaction, such as providing loop invariants. User interaction, however, requires highly specialized expert knowledge, which is unreasonable for typical software developers~\cite{BFL+11}.

While most emphasis in the formal methods community is put on the scalability of the verification phase, scalability of the specification phase has only gained momentum recently~\cite{baumann2012lessons,beckert2016deductive}. Applying \emph{Design by Contract}, a methodology in the context of deductive verification based on Hoare triples~\cite{M88}, developers specify the behavior of methods with contracts comprising first-order preconditions and postconditions. Contracts are typically annotations in source code close to the implementation~\cite{BFL+11,C:NFM11,M88}. The rationale is whenever a caller guarantees the precondition, the method guarantees its postcondition. A caller can then reason about its correctness with respect to its own contract. 

However, even when applying Design by Contract in the industry, software developers typically only assure the quality of a small subset of the code base by writing or modifying contracts alongside their implementation. These developers are either not verifying their implementation at all or are at least never in a position to verify that their contracts are sufficient for all callers, which is often done by a different team focusing on the overall quality assurance. Consequently, a considerable amount of insufficient contracts may emerge that must be adapted subsequently. Providing strong enough contracts is error-prone and tedious~\cite{baumann2012lessons} and, thus, cost-intensive. 

We argue that, to become applicable in industrial software development, deductive verification including the formal specification phase must be supported by easy-to-use tools with a high degree of automation, such that typical software developers can already contribute significantly to the implementation's quality. To identify hurdles and challenges, experience reports are indispensible. Our observations are based on software written in objected-oriented languages and specified with contracts. In particular, we draw our experiences from a real-world case study, where we specify and verify parts of OpenJDK's Collections-API with the Java Modeling language (JML)~\cite{LC06}. For the verification phase, we use the state-of-the-art verifier KeY version 2.6.1~\cite{ahrendt2016deductive}, an interactive theorem prover with a high degree of automation and a large community.
Our long-term vision is to facilitate the development of specification and implementation in concert for  everyday software developers. Our work contributes to this goal by investigating usability and applicability of KeY from a user's persepective, as most studies in research about specification and verification with KeY are indeed conducted by KeY developers themselves (\eg~\cite{baumann2012lessons,beckert2017proving,de2015openjdk}). 

Our real-world case study exhibits that formal specification and automatic verification of APIs is in principle feasible. It also shows that significant experience is necessary and, accordingly, that current verification technology is not yet suitable for less experienced developers. In this sense, we believe that a community effort is indispensable to specify widely used APIs such as OpenJDK, which would help users tremendously to verify their own implementation against it. In summary, our main contributions are the following.

\begin{itemize}
\item We describe our experiences on specifying parts of OpenJDK's Collection-API with JML based on its existing informal specification (\ie JavaDoc) and verify it using the program verifier KeY in version 2.6.1.
\item We distribute OpenJDK's source code together with our formal specification online with the goal to extend it continuously.
%\item We show that
\item We discuss which characteristics impair automation in the verification phase and give examples.
\item We encounter an incidence where KeY proves unsound behavior due to Java's under-specification of the maximal array length.
\end{itemize}

\noindent The paper is structured as follows. In Section~\ref{sec:contracts}, we provide the necessary background on contract-based deductive verification of Java programs. In Section~\ref{sec:specification}, we explain how we specified parts of OpenJDK's Collections-API and discuss challenges. In Section~\ref{sec:experiences}, we provide a detailed experience report in specifying and verifying  Java programs with KeY from a developer's perspective. We discuss related work in Section~\ref{sec:related} and conclude this work  in Section~\ref{sec:conclusion}.

%% file: content/contracts.tex
\section{Formal Verification of Java Programs}
%From informal to formal...
\label{sec:contracts}

Quality-assurance techniques, such as code reviews, testing, and formal methods, are critical for safety-related software. A prominent approach that is part of many modern programming languages is to use \emph{assertions}~\cite{F67}, which are propositional formulas that should always be satisfied at given locations of method execution. A generalization of assertions is the design-by-contract paradigm~\cite{LG86,M88,M92}, which comes with dedicated specification languages supporting flavors of first-order logic. Contracts decorate methods of object-oriented code with \emph{preconditions} $\psi$ and \emph{postconditions} $\phi$, and classes with \emph{class invariants}. Preconditions describe what a method can assume and must be provided by callers of that method. Postconditions describe what a method must guarantee if its preconditions are fulfilled. Invariants must always hold (\ie before and after method execution). While assertions are typically checked at runtime, \emph{deductive reasoning}, as applyed by theorem provers, is used to verify source code statically~\cite{Sch01}. A program together with its specification (\ie contracts) is translated into a flavor of dynamic logic. The resulting logical formula for a method $m$ is then proved to always hold by systematically applying inference rules to it. 

%\subsection{Java Modeling Language}
For specification, there exist numerous languages with support for contracts, such as Eiffel~\cite{M92}, Spec\#~\cite{BFL+11}, and the Java Modeling Language (JML)~\cite{LC06}. For the purpose of this paper, we focus on JML, a contract-supporting extension of Java for specification. The reason is that Java is far  more widespread than the other languages and highly applied in industry and research.
In \ref{fig:jml-example},
\begin{lstlisting}[float=t!,label=fig:jml-example,caption={[Test]An Account Implementation with Contracts in JML}]
class Account {
	public final static int DAILY_LIMIT = 1000;
	public int balance;
	public int withdraw;
	/*@ |\requires| withdraw < DAILY_LIMIT, amount !=0;
	  @ |\ensures| (|\result| == (|\old|(withdraw) - amount < DAILY_LIMIT))
	  @ && (|\result| ==> withdraw == |\old|(withdraw) - amount)
	  @ && (|\result| ==> balance == |\old|(balance) + amount);
	  @ |\assignable| withdraw, balance;
	  @/
	boolean update(int amount) {...}
}
\end{lstlisting}
we give an example of a concrete contract written in JML of a method \code{update} in a class \code{Account}. Method \code{update} manages the transfer of money from and to respective accounts. The precondition is denoted by keyword \code{requires} and states that callers of method \code{update} must ensure that (a) input \code{amount} is not equal to 0 and that (b) value of field \code{withdraw} is less than the value of the field \code{DAILY\_LIMIT}. The postcondition is denoted by keyword \code{ensures} and states that method \code{update} guarantees to update withdraw and balance, whenever the daily limit is still not reached. In a postcondition, keyword \code{old} refers to the state of the expression before method execution and keyword \code{result} represents the return value. Moreover, the framing clause is denoted by keyword \code{assignable} and describes which part of the memory is allowed to be changed by the method's implementation. If all classes are known, the framing clause is syntactic sugar and may also be expressed in the postcondition by using \code{old(v) == v} for all locations \code{v} that should remain unmodified.

As can be seen in~\ref{fig:jml-example}, JML provides means to specify the intended behavior of a method close to the implementation. Besides method contracts, JML also allows to specify invariants for fields, loops with \emph{loop invariants}, and even blocks (\ie scoped statements in curly braces) with \emph{block contracts} inside an implementation. Additional specifications are often necessary to increase automation and decrease interaction in the verification phase.

%#4 java specifics to consider (integers, arrays, informal verification)

%#5 KeY: supports java card, automation, bezug zu developers, parameters

%\subsection{The Deductive Verification System KeY}
To investigate the difficulties of specifying and verifying source code from a users perspective, we focus on KeY~\cite{ahrendt2016deductive}, an interactive theorem prover for JML-specified Java programs with a highly active community. KeY is a frequently used verification system for JML with the goal to bridge the gap between research and industry. In particular, KeY translates specification and implementation to an extension of dynamic logic called \emph{Java Dynamic Logic}~\cite{ahrendt2016deductive}. Resulting proof obligations are processed step-wise based on a combination of symbolic execution~\cite{ahrendt2016deductive} and weakest precondition calculus~\cite{H:JMLC03} and are either closed automatically or remain open to be manually inspected by a user. 

The goal of \emph{Design by Contract} is to write implementations together with their contracts in concert~\cite{M88}. Despite the fact that most software developers are non-experts in formal verification, they typically know all requirements that are important for the code they introduce or modify and, thus, should be supported by the verification system in use to write concise and comprehensible formal specifications. Moreover, to ease the process of specification and increase applicability of formal verification even into the realm of mainstream software development, the verifying tool chain has to provide a high degree of automation, which is in-line with the Spec\# experience~\cite{BFL+11}. Besides the fact that KeY was initially designed to be used interactively~\cite{ahrendt2016deductive}, it provides numerous means to automate the verification process. For instance, KeY applies sophisticated built-in strategies  to find proofs automatically. Developers may even define and add their own strategies. Moreover, KeY offers a considerable amount of parameters that  control how the automatic proof search behaves. Setting the right parameters purposefully requires  expertise, but also allows a user to decrease the verification effort significantly~\cite{KTPS:ITP18}.

%% file: content/specifying-collection.tex
\section{Contract-Based Specification of the Collection-API}
\label{sec:specification}

A challenge to address in formal verification is to formally specify a given implementation sufficiently, such that it can be verified automatically -- in particular when performed by less experienced developers. 
To gain experiences in this regard and identify hurdles and challenges for typical software developers, we carried out a real-world case study, for which we decided to specify parts of OpenJDK~6. Reasons to use OpenJDK are manifold. One of our goals is to specify and verify widespread and highly applied real-world software. Whereas building software from scratch was therefore not an option, OpenJDK qualifies for these requirements and is also open source. Moreover, OpenJDK is free to distribute, even when the source code is altered (\eg adding JML contracts). Oracle's JDK disqualifies for the very same reason. Another point is that reconstructing a developers intention to develop a formal specification is difficult (\ie in case of legacy systems). However, OpenJDK already provides a comprehensive informal specification in its JavaDoc, which eases the development of a formal specification.

\subsection{Specifying the Java Collection-API}

For specifying methods in Java programs with contracts, we use the Java Modeling Language (JML). In \ref{lst:infSpecCopyOf},
%\begin{lrbox}{\mybox}
\begin{lstlisting}[float=t!,caption={JavaDoc and Signature of \code{Arrays.copyOf()}},label={lst:infSpecCopyOf}]
/**
Copies the specified array, truncating or padding with nulls (if necessary) so the copy 
has the specified length. For all indices  that are valid in both the original array and 
the copy, the two arrays will contain identical values. For any indices that are valid in 
the copy but not the original, the copy will contain null. Such indices will exist if and 
only if the specified length is greater than that of the original array. The resulting 
array is of exactly the same class as the original array.

@param original the array to be copied
@param newLength the length of the copy to be returned
@return a copy of the original array, truncated or padded with nulls 
to obtain the specified length
@throws NegativeArraySizeException if <tt>newLength</tt> is negative
@throws NullPointerException if <tt>original</tt> is null
@since 1.6
*/
public static Object[] copyOf(Object[] original, int newLength)|\modsep|
/*@
  @ |\public| |\exceptionalbehavior|
  @ |\requires| newLength < 0;
  @ |\signals| (NegativeArraySizeException e) true;
  @
  @ |\also| |\public| |\exceptionalbehavior|
  @ |\requires| original == null;
  @ |\signals| (NullPointerException e) true;
  @
  @ |\also| |\public| |\normalbehavior|
  @ |\requires| original != null && newLength >= 0;
  @ |\ensures| |\result| != null && |\fresh|(|\result|) && |\result| != original
  @		&& |\typeof|(\result) == |\typeof|(original) && |\result|.length == newLength;
  @ |\ensures| (|\foralljml| int i; 0 <= i && i < |\result|.length &&
  @		i < original.length; original[i] == |\result|[i]);
  @ |\ensures| (|\foralljml| int i; original.length <= i && 
  @		i < newLength; |\result|[i] == null);
  @*/
public static /*@ nullable pure@*/ Object[] copyOf(/*@ nullable @*/ 
     Object[] original, int newLength)
\end{lstlisting}
%\end{lrbox}
%\scalebox{0.8}{\scalebox{0.8}{\usebox{\mybox}}
we exemplify the process of specifying a method based on an informal specification on the method \code{copyOf} for class \code{Array}.  At the top we depict the informal specification provided in the JavaDoc comments. In essence, the informal specification covers the following aspects if method \code{copyOf} successfully terminates.
\begin{enumerate}
\item The result is a new array with given length \code{newLength}.
\item If \code{newLength} is less than the length of \code{original}, the resulting array is truncated.
\item If \code{newLength} is greater than the length of \code{original}, the resulting array is padded with \code{null}-elements.
\item Values of all indices that are valid in \code{original} and the resulting array are identical.
\item Array \code{original} and the resulting array have the same type. 
\end{enumerate}
Below the JavaDoc comment, we present the JML contract that we derived for that informal specification. In concert with the informal specification, the contract comprises three specification cases confined by keyword \code{also}; in two cases, exceptions are thrown when \code{newLength} is negative or array \code{original} equals \code{null}. The third specification case depicts the intended behavior as explained before.

The \emph{Java Collection-API} provides an architecture to temporarily store and manipulate a group of objects. The interface \code{java.util.Collection} is the foundation for numerous data structures and is, for instance, implemented by \code{java.util.List} and \code{java.util.Set}. In the specification process, we concentrated on a small number of methods of the collection interface that we wanted to specify and verify. Some prominent methods of the collection interface we focused on are the following.
\begin{itemize}
\item \code{int size()}: returns the number of objects.
\item \code{boolean isEmpty()}: informs whether the number of objects is zero.
\item \code{boolean contrains(Object)}: informs whether the collection holds a specific object.
\item \code{boolean add(Object)}: adds an object and returns \code{true} if the collection changed.
\item \code{boolean remove(Object)}: removes an object and returns \code{true} if the collection changed.
\item \code{void clear()}: removes all objects.
\item \dots
\end{itemize}

Here, we follow a bottom-up approach in specifying the implementation; we first specify and verify less complex methods associated with a small call stack when executed. The rationale is that sufficient contracts to automatically verify such a method should be easier to derive, as only a few dependencies to called methods exist. Subsequently, we can specify and verify more complex methods associated with larger call stacks when executed by relying on specifications of called methods we derived before. Moreover, strong enough postconditions are in some cases easier to identfy compared to a top-down approach, as the postcondition of a caller depends to a great extend on the postconditions of called methods. However, as discussed in more detail elsewhere~\cite{baumann2012lessons,beckert2016deductive}, neither pure bottom-up nor pure top-down approaches are always superior in general.

\subsection{Lessons Learned in Formalizing Parts of OpenJDK}
In this section, we elaborate on our process and gained experiences of specifying parts of the Collections-API with JML. Our specifications can be found online and we invite other researchers to contribute to that repository and extend it in the future.\footnote{\url{http://github.com/AlexanderKnueppel/OpenJDKwithJML/tree/F-IDE18}}

\subsubsection*{Behavioral Subtyping.}
In the presence of subtyping, contracts of a type and its subtypes should follow  behavoral subtyping~\cite{LW:TOPLAS94}. If \code{T} and \code{S} are both types where \code{S} is subtype of \code{T}, behavioral subtyping states that \code{T} can be replaced by \code{S} in any scenario where \code{T} is used without distorting a program's behavior. For deductive verification, this means that we can always use \code{S.m()} instead of  \code{T.m()}  and that their contracts must therefore be in a compatible relation (\eg preconditions of \code{S.m()} can only be strengthened, whereas postconditions can only be weakened). The reason to follow this principle is that behavioral subtyping enables modular reasoning~\cite{leavens2006behavioral}, as, even in case of dynamic dispatching, the supertype can be used. In particular, JML features a particular instance of behavioral subtyping called \emph{specification inheritance}, which is enforced by KeY. Specification cases of an overriding method are conjoined with the supertype's specification cases by employing keyword \code{also}.

The collection interface is highly generic and refers in its informal specification (\ie JavaDoc) to numerous properties that subclasses may either establish or not. An example for a property is whether a subclass allows storing of duplicates. To follow behavioral subtyping, a considerable amount of methods are directly specified by us in the collection interface and inherited by subclasses. Thus, we need to take these properties into account when generally specifying a method. That is why we started to introduce \emph{model fields} in the collection interface, which allows us to parameterize contracts (\ie parameters are used in contracts and instantiated in concrete subclasses individually). Model fields can only be used in JML-annotations and represent states that are evaluated in the verification process.

In \ref{lst:model}, 
\begin{lstlisting}[float=t!,caption={Model Fields of the Java Collections Framework},label={lst:model}]
  /*@
  @	|\public| |\modeljml| boolean supportsDuplicates;
  @ |\public| |\modeljml| boolean supportsNull;
  @ |\public| |\modeljml| boolean isOrdered;
  @
  @ |\public| |\modeljml| instance \bigint collectionSize;
  @ |\public| |\modeljml| instance nullable Object[] elements;
  @
  @ |\public| |\modeljml| instance \locset changeable;
  @*/
\end{lstlisting}
we illustrate all model fields that we use for the Collection-API. For example, some implementations do not allow to hold duplicates (\eg \code{Set}), which is why we specified a \code{boolean} variable named \code{supportsDuplicates} for this purpose. Classes that do not allow to hold duplicates instantiate this field with \code{false}. 
% specification on the interface
%Die im vorherigen Unterkapitel erstellten Felder existieren jetzt zwar, haben jedoch noch keine Werte, da diese in dem Collection-Interface nicht bekannt sein konnen. Die Belegung mit Werten wird von jeder Klasse separat vorgenommen.
%\paragraph{Solution}

\subsubsection*{Informal Specification is Often Imprecise.}
OpenJDK already provides a comprehensive informal specification for most methods in form of JavaDoc comments (cf.\ \ref{lst:infSpecCopyOf}). An inherited problem of an informal specification is, however, its imprecise nature.  A consequence is that imprecision impedes the direct translation to a formal specification. Moreover, methods may depend on numerous other methods and in the process of software evolution specifications can become outdated -- especially if they are not verifiable. 

We encountered that, in some cases, implicit behavior had to be made explicit, particularly for private methods. For example, the insufficient informal specification of \code{ArrayList.fastRemove} is the following: \emph{Private remove method that skips bounds checking and does not return the value removed}.
We also found cases where an informal specification was not reasonable and had to be ignored. An example is the \code{size()} method, where the informal specification states that \emph{whenever the current size of the collection is greater than  \code{Integer.MAX\_VALUE}, value \code{Integer.MAX\_VALUE} has to be returned}. This, however, is not compliant with the implementation. Based on our experience,  informal specifications are often to a great extent incomplete and erroneous. Defects even remain incognito for years or decades, which consequently means that a direct translation from an informal specification is often impractical and has to be taken with caution.

%\paragraph{Solution}

\subsubsection*{Missing Tool Support for Contract-Based Specification.}
Ideally, the verification system provides means to support the specification phase as well as the verification phase. For instance, early feedback is crucial to prevent unnecessary iterations in specification and re-verification, which is time consuming and error prone. However, following a modular approach, most currently active verification systems only focus on the method under verification and do not consider callers of that method. 
For instance, when a user specifies a method, some feedback on whether the contract is sufficient for callers would be helpful. Moreover, completely specifying a large code base is unrealistic. Hence, when given a set of prime methods that need to be verified, a small set of additional specified methods may help to decrease the verification effort significantly. In the future, identifying those methods becomes crucial to bridge the gap between research and industry.

To ease the process for industrial software developers, the hurdles of contract-based specification must decrease through better tool support as well. Although a developer must know how to specify an implementation with contracts, there is still room for improvements. For instance, there exist an enormous amount of research on automatic inference of loop invariants or generating specification cases in case of \emph{trivial} implementations, but its practically remains to be investigated. Especially under change, when specification or implementation are easily violated, automated reasoning and feedback for resolving such violations (\eg suggesting fixes) are needed to prevent the deployment of bugs or costly re-verification.

%\paragraph{Solution}

%% file: content/experiences.tex
\section{Contract-Based Verification with KeY}
\label{sec:experiences}
In the following, we exhibit our experiences and results of verifying parts of OpenJDK~6 with KeY 2.6.1 (cf.\ Section~\ref{sec:results}) and discuss challenges that software developers face when applying  deductive verification automatically (cf.\ Section~\ref{misunderstandings}). 
%In the verification pahase with KeY

\subsection{Experiences Drawn from Verifying OpenJDK}
\label{sec:results}

%\subsubsection*{Issue: Integer Overflow}

\subsubsection*{Invariant Checking in Constructors}

In \ref{lst:constr},
\begin{lstlisting}[float=t!,caption={Specification of \code{ArrayList(int)}},label={lst:constr},breaklines=true]
/*@ elementData != null;
@ |\invariant| |\typeof|(elementData) == |\typeof|(Object[]);
@*/
private transient Object[] /*@ |\nullablejml| @*/ elementData;
...
/*@ |\public| |\normalbehavior|
  @ |\requires| initialCapacity >= 0;
  @ |\ensures| elementData.length == initialCapacity;
  @ |\assignable| elementData;
  @ |\also|
  @ |\public| |\exceptionalbehavior|
  @ |\requires| initialCapacity < 0;
  @ |\signalsonly| IllegalArgumentException;
  @ |\signals| (IllegalArgumentException e) true;
  @*/
public ArrayList(int initialCapacity) {
	super();
	//this.elementData = new Object[0]; //resolves  the problem
	if (initialCapacity < 0)
		throw new IllegalArgumentException(" Illegal Capacity :  "+
		initialCapacity);
	this.elementData = new Object[initialCapacity];
}
\end{lstlisting}
we illustrate the normal and exceptional specification of \code{ArrayList}'s constructor. Whereas the normal behavior is verifiable, the exceptional behavior is indeed not. The reason is that if an exception is thrown, \code{this.elementData} will not be instantiated (\ie remains \code{null}). However, this contradicts the invariant stating that  \code{this.elementData} must not be equal to \code{null}. The question here is whether the invariant should be checked even if object construction fails. We postpone the question to Section~\ref{misunderstandings}.

\paragraph{\emph{Resolution:}} To resolve this issue and prove the constructor's correctness in the exceptional case, we can modify the implementation and instantiate \code{this.elementData} with zero elements before an exception can be thrown. However, checking the invariant  in this specific case when object construction fails seems to be unnecessary. Providing means -- preferably though an extended specification -- to hinder invariant checking in specific cases would be desirable. 

Indeed, there exists the possibility to explicitly declare the constructor as a helper method using the keyword \code{helper} in front of the method's name~\cite{LM:ICSE07}. Helper methods exclude checking invariants in their pre- and postconditions. In this specific example, however, using \code{helper} is not an appropriate  solution. First, we may only want to hinder checking invariants in particular specification cases (\ie the exceptional behavior). Second, the reference manual for JML states that helper methods and constructors need to be declared as \code{private}. 

\subsubsection*{Pure Methods without Specification}
A different problem we encountered by specifying \code{List.indexOf(Object)} was that the method could not be verified automatically due to the usage of \code{Object.equals(Object)}. \code{List.indexOf(Object)} returns the lowest index in a list where its element equals the input object or \code{-1} if there is no such element. In its informal specification, equality is based on the equals method of the respective object. \code{Object.equals()} is used in the specification  as well as the implementation, but does not have a specification itself. Thus, each call of it is replaced by its implementation in the verification process. If we replace each call to \code{Object.equals()} in the implementation with \code{==} manually (which is generally against the intention of \code{List.indexOf(Object)}) the method becomes verifiable. Replacing \code{Object.equals(Object)} in the specification with \code{==}, however, does not help. 

\paragraph{\emph{Resolution:}} Our assumption is that pure methods called in the specification are treated differently than in the implementation. In both cases, method calls where no contract exists should be replaced by its implementation, which apparently only happened in the specification.

%\subsubsection*{Method getClass() and Native Methods.}
%Methods \code{Arrays.copyOf(Object[], int)} and \code{Arrays.copyOfRange(Object[], int, int)} have been specified but not verified. The reason is that both rely on method \code{getClass()} for which we did not specify a contract. The reason is that specifying method \code{getClass()} would only delegate the problem; method \code{getClass()} calls method \code{Arrays.copyOf(Object[], int, Class)}, which internally uses field \code{.class} of an object and the native method \code{System.arraycopy(...)}. We found out that field \code{.class} cannot be parsed with KeY 2.6.1 and for native methods we do not have an implementation.
%
%\paragraph{Resolution:} It is possible to provide contracts for called methods without verifying them. Those contracts are then used as a replacement for method calls.  However, we have to \emph{believe} that the hidden implementation adheres to its specification.

\subsubsection*{Underspecification of the Java Semantics (Maximal Array Length)}
We experienced an issue with method \code{ArrayList.toArray()}, when we tried to verify it. KeY does not check whether the length of an array is \emph{too big}. In particular, there is no maximal array length specified in the Java language specification and, thus, KeY, as well as other tools, have no obligation to provide a check for it. Problematic is that developers of virtual machines can set their own maximal array length, which is often around \code{Integer.MAX\_VALUE - 4} (\ie specific bytes are reserved for header information), but also may change from version to version. A typical user would expect the maximal array length to be equal to the maximal integer value (\ie according to the informal specification). 

One goal of design by contract is to render \emph{defensive programming} needless (\ie checking in the implementation that the input is in the right range), as it can be considered redundant to a formal specification. Moreover, formal verification is a means to not only ensure safety properties but also to prevent security-related problems. Both objectives are invalidated by this issue. In practice, an attacker could exploit this problem by providing an array as input of approximately 2GB in size, which would consequently crash the program. 

\paragraph{\emph{Resolution:}} KeY and other tools could offer the possibility to dynamically set the maximum length of an array in its front-end (\ie this option would behave like a global invariant for all arrays). Otherwise, users of KeY and possibly other verification systems must be aware of this problem and provide maximal array lengths themselves.

\subsubsection*{Parameterization}
A different challenge we faced is parameterization. In the front-end of KeY, a user can select numerous options to control how KeY should try to automatically solve the current verification task. Some options are trivial, such as \emph{method call treatment} (\ie whether the contract of a called method is used or the implementation is inlined). Other options, however, are unclear to typical users and require deep knowledge about KeY's underlying theory and solving procedure. This is particularly a problem, as various options and specific combinations thereof strongly influence provability and verification effort. Additionally, the number of options rather increases with each new version~\cite{KTPS:ITP18}. Finding the right settings for the current verification task is therefore indispensable. For instance, we faced a problem during the verification of method \code{ArrayList.add(Object)}, where it could only be verified if we ignore integer overflows (\ie there exists a specific parameter to either ignore or check for overflows, or directly rely on the Java semantics for integers). This problem is connected with the unspecified length (\ie field \code{size}) of a collection. 

\paragraph{\emph{Resolution:}} One ad-hoc solution is to perform trial and error, which is exceptionally time consuming, but consequently leads to a set of prime configurations for different verification tasks. A more sophisticated approach was proposed in a complementing study~\cite{KTPS:ITP18}, where we conducted an empirical study on the influence of  parameters with respect to provability and verification effort. Such experiments help to derive a guideline that a user can follow to identify when to use which option. One can even consider setting options automatically, when the implementation is easy enough to process by an algorithm, or at least extended tool support by integrating a recommendation system.

\newcommand\wrong[1]{\bfseries\sffamily #1}
\newcommand\proven[1]{\bfseries\sffamily #1}
\newcommand\problematic[1]{\bfseries\sffamily #1}

\subsubsection*{Summary} 

In Table~\ref{tab:results},
\begin{table}[t]
\centering
  \begin{tabular}{p{2cm}p{4cm}cc}
  	Class  & Method & \# spec. cases& Proof result \\ \hline	
	ArrayList & add(Object) & 6 & \wrong{Overflow problem} \\
	ArrayList & \textit{constructor}() & 1 & \proven{Proven} \\
	ArrayList & \textit{constructor}(int) & 2 & \wrong{Invariant problem} \\
	ArrayList & clear() & 2 & \proven{Proven}\\
	ArrayList & contains(Object) & 3 & \proven{Proven} \\
	ArrayList & ensureCapacity(Object) & 3 & \proven{Proven}\\
	ArrayList & fastRemove(int) & 1 & \proven{Proven}\\
	ArrayList & indexOf(Object) & 1 & \problematic{Works with ==, not with equals}\\
	ArrayList & isEmpty() & 1 & \proven{Proven}\\
	ArrayList & outOfBoundsMsg(int) & 1 & \proven{Proven}\\
	ArrayList & rangeCheck(int) & 1 & \proven{Proven}\\
	ArrayList & rangeCheckForAdd(int) & 1 & \proven{Proven}\\
	ArrayList & remove(Object) & 3 & \problematic{Works with ==, not with equals}\\
	ArrayList & size() & 1 & \proven{Proven}\\
	ArrayList & toArray() & 1 & \parbox[t]{6cm}{\centering \wrong{Proven - but Array instantiation problem}.}\\
	ArrayList & trimToSize() & 1 & \proven{Proven}\\\hline
	Arrays & copyOf(Object[], int) & 3 & \problematic{Uses getClass()} \\
	Arrays & copyOfRange(Object[], int, int) & 4 & \problematic{Uses getClass()} \\\hline
	Math & abs(int) & 1 & \proven{Proven}\\
	Math & max(int, int) & 1 & \proven{Proven}\\
	Math & min(int, int) & 1 & \proven{Proven}\\
  \end{tabular}
  \caption{Specified Methods and Proof Results of our Real-World Case Study}\label{tab:results}
\end{table}
we summarize all initially specified methods and respective proof results. On average, 71\% of all methods could be verified automatically.  We wrote a total of 175 lines of JML specification in the presented classes (excluding some inherited specification on the interface level) over the course of four person months. 
In this process, we  encountered numerous obstacles, which we either resolved (cf.\ Section~\ref{sec:results}) or which may require additional effort by the community. Namely, these obstacles were the problem of integer overflow, the problem of checking invariants when object construction fails, the problem with \code{equals()}, the problem with too big arrays, and a problem with \code{getClass()}, which internally depends on the \code{.class} field. \code{.class}, however, was not parsable by KeY. 
Based on these results, we conclude that, in principle, verification of real library code is practically in reach, but the specification process is extremely time-consuming.

A cumulated overview of lines of Java code, lines of JavaDoc, and written lines of JML specification is depicted in Table~\ref{tab:summary}.
{
\renewcommand{\arraystretch}{1.15}
\begin{table}[t]
\centering
\begin{tabular}{l*{3}{c}|ccccc}
&  &  & & \multicolumn{5}{ c }{\textbf{Lines of JML}} \\\cline{5-9}
Classes             & Methods & Java lines & JavaDoc lines & Invariant & Requires  & Ensures & Other & \textbf{Total}\\
\hline
ArrayList & 16&  96 & 57 & 11 & 30 & 29 & 46 & \textbf{116}\\
Arrays    & 2& 12 & 35 & 0 &  10 & 14 &  24 & \textbf{48}  \\
Math      & 3& 6 &  23 &  0 & 0   & 7 & 4  & \textbf{11} \\\hline
\textbf{Total} & \textbf{21} & \textbf{116} & \textbf{115}  & \textbf{11} &  \textbf{40} & \textbf{50} &  \textbf{74} & \textbf{175}\\
\end{tabular}
  
  \caption{Statistics on the Specified Part of OpenJDK}\label{tab:summary}
\end{table}
}
Field \emph{Other} refers to any line that is neither associated with an invariant, a precondition, nor a postcondition (\eg keyword \code{nullable} or exception handling through \code{signals}). The specification contains 50\% more lines than the actual implementation.

\subsection{Misunderstandings from a User's perspective}
\label{misunderstandings}
Occasionally, KeY acted differently than we expected. In the following, we want to highlight some of these occurrences in more detail to give tool builders feedback with respect to practical application and to help users who encounter the same situations.

\subsection*{Array Instantiation - Nullable}

In \ref{lst:minBeispArrayInst},
\begin{lstlisting}[float=t!,caption={Two Identical Methods with Almost Identical Specifications},label={lst:minBeispArrayInst},breaklines=true]
/*@ |\public| |\normalbehavior|
  @ |\requires| arrSize >= 0;
  @ |\ensures| |\result| != null;
  @ |\ensures| |\result|.length == arrSize;
  @*/
private static Object[] cArrayNotWorking(int arrSize){
   return new Object[arrSize];
}

/*@ |\public| |\normalbehavior|
  @ |\requires| arrSize >= 0;
  @ |\ensures| |\result| != null;
  @ |\ensures| |\result|.length == arrSize;
  @*/
private static /*@ |\nullablejml| @*/ Object[] cArrayWorking(int arrSize){
   return new Object[arrSize];
}
\end{lstlisting}
we depict two identically implemented methods. The specification of the second method is extended by keyword \code{nullable} for its return value. The second method is verifiable, whereas the first method is not. The reason is that keyword \code{nullable} does not only affect the reference object itself but also all of its elements. Indeed, the default behavior in JML enforces that all reference types must be non-null and, thus, elements of the instantiated array must also be non-null. In case of \code{arrSize\textgreater 0} and non-primitive data, however, both methods fill the array with null elements according to Java's default behavior for initializing reference types. 

This \emph{trivial} mistake was hard to spot for us in the first few attempts. The open proof goal in KeY and he symbolic execution debugger did also not provide enough information to resolve this issue. In our experience,  developing techniques to infer trivial specifications from the code to at least give suggestions and feedback to users about what may be missing is crucial in the future to speed up the specification process.

%\subsection*{Signals in Multiple Specification Cases}
%
%In \ref{lst:minBeispExBe},
%\begin{lstlisting}[float=t!,caption={Example of Two Overlapping Exceptional Specification Cases},label={lst:minBeispExBe},breaklines=true]
%/*@ |\public| |\exceptionalbehavior|
%  @ |\requires| original == null;
%  @ |\signals| (NullPointerException e) true;
%  @ |\also|
%  @ |\public| |\exceptionalbehavior|
%  @ |\requires| newLength < 0;
%  @ |\signals| (NegativeArraySizeException e) true;
%  @*/
%public static /*@ |\pure| @*/ void method(/*@ |\nullablejml| @*/ Object[] original, int newLength){
%	if(original == null)
%		throw new NullPointerException();
%	if(newLength < 0)
%		throw new NegativeArraySizeException();
%}
%\end{lstlisting}
%we illustrate two overlapping specification cases. Whenever \code{original == null} and \code{newLength\textless 0}, both specification cases are valid and, consequently, one can assume that both signals trigger the respective exception. That these specifications are valid is based on the semantic of \code{also}, where preconditions are conjoined with an or-relationship and if one of the two exceptions is thrown, it implies its respective postcondition. The rationale is to give the implementation more freedom. The implementation in this case provides an order of throwing exceptions; if both preconditions hold, the \code{NullPointerException} always comes first. 

\subsection*{Invariants in Constructors}

\begin{lstlisting}[float=t!,caption={Example of a Constructor that Saves the Reference in a Static Field},label={lst:constWithExc},breaklines=true]
public class ExceptionalConstructor {
	public static final List<ExceptionalConstructor> created = new ArrayList<>();
	public boolean initialized;

	public ExceptionalConstructor(boolean throwing) {
		super();
		created.add(this);
		if (throwing) {
			throw new RuntimeException();
		}
		initialized = true;
	}
}
\end{lstlisting}
As mentioned before, invariants are checked in constructors even when object construction fails. Despite the fact that this behavior does not seem to be intuitive, we depict a minimal example in \ref{lst:constWithExc}, where we believe that this behavior is indeed reasonable. Although the custom constructor of class \code{ExceptionalConstructor} may fail, its reference is saved in a static field. In this case, the constructor of \code{ExceptionalConstructor} would not return the reference, but saving the reference is possible, because, according to Java's semantics, the object is created even before the constructor and its superconstructors are executed. The question is, again, whether all invariants must hold in any given scenario. 

A solution would be to provide a new keyword for excluding invariants in specific cases or to use a boolean \emph{ghost field} (similar to a model field) which is set to true once the invariant is established. The invariant can then be rewritten as an implication (\ie \code{\textbackslash *@ invariant ghost ==\textgreater \, arr != null; @*\textbackslash}).

\subsection*{Array Access - Data Types}

\begin{lstlisting}[float=t!,caption={Minimal Example of an Array Access},label={lst:minArrDatTyp},breaklines=true]
public class ArrayStoreValid{
    /*@ |\invariant| arr != null;
      @ |\invariant| arr.length == 1;
      @*/
    private transient /*@ |\nullablejml| @*/ Object arr[];
	
    /*@ |\public| |\normalbehavior|
      @ |\ensures| true;
      @*/
    public void set(Object o) {
        arr[0] = o;
    }
}
\end{lstlisting}
Given the specification of \code{ArrayStoreValid.set(Object)} illustrated in \ref{lst:minArrDatTyp}, this method appears to be easily verifiable. However, KeY fails to verify this method and provides an \code{ArrayStoreException}. The reason is based on array \code{arr}'s type; \code{arr} can take on any subtype of \code{Object[]}. Moreover, \code{arr} can hold elements of any subtype of \code{Object}. Both types may be incompatible, which is recognized by KeY.
This issue can be resolved by adding the invariant
\code{\textbackslash *@ invariant \textbackslash typeof(arr) == \textbackslash type(Object[]);@*\textbackslash} to explicitly inform KeY that array \code{arr} will always be of type \code{Object[]}. However, again, resolving such issues should be supported by additional tooling during the specification phase.

%% file: content/related.tex
\section{Related Work}
\label{sec:related}

In the following, we discuss differences to related research that also focuses on specification and verification of software systems with contracts.

%\paragraph{Specification with Behavioral Contracts}
A survey on different languages for behavioral contracts was done by Hatcliff et
al.\ \cite{HLL+12}. Besides JML, there exist alternatives for specifying Java source code, such as C4J~\cite{BM13} or Contract4J~\cite{wampler2006contract4j}. Other examples for languages with support for contract are Eiffel~\cite{M92} and Spec\#~\cite{BFL+11}. We consider our results to be generally applicable to other languages, as the specification and semantics of those contract languages is similar to JML.

%\paragraph{Verification of Object-oriented Source Code}
For the purpose of this paper, we chose KeY 2.6.1~\cite{ahrendt2016deductive} as the primary verification system. 
There are a number of verifiers that have been used in substantial verification efforts. Dafny was employed in IronClad and IronFleet~\cite{hawblitzel2015ironfleet,hawblitzel2014ironclad}, Autoproof has been used in the development of a verified Eiffel library~\cite{furia2017autoproof,polikarpova2015fully}, and F* has been used in Microsoft's project Everest~\cite{bhargavan2017everest}.
For object-oriented programming in general, there exist a number of alternatives. The KIV system~\cite{ernst2015kiv} can be employed for the development of safety-related software, is also based on a dynamic logic, and primarily focuses on strong proof support. ESC/Java2~\cite{cok2004esc} is a static checker that finds common runtime-errors in JML-specified programs. Other systems are Krakatoa~\cite{FM07} for Java programs and Jesse~\cite{marche2012jessie} for C programs, which are both based on the Why platform for deductive reasoning~\cite{why}. Both systems require high expertise, as proof scripts are manually written by users and also appear to be highly brittle~\cite{TSKA:VAST11}. Typically, verification of imperative languages following the design-by-contract paradigm are based on first-order logic, such as KeY that is based on Java dynamic logic. Examples of interactive theorem provers for higher-order logic are Coq~\cite{BC04}, Isabelle/HOL~\cite{NWP02}, and PVS~\cite{ORR+:CAV96}. Nevertheless, for our real-world case study we chose KeY, because none of these systems is directly designed to support verification of mainstream programming languages by mainstream software developers.

%\paragraph{Real-life case studies}
Despite being a research topic for decades, formal methods are still not widely applied by industrial software developers. An often in research overlooked challenge is the difficulty of formally specifying real-world software. Thus, only a few publications exist that either discuss necessities for formal specifications to become widely applicable or discuss real-world case studies. Beckert et al.\ \cite{beckert2016deductive} discuss strategies and requirements for contract-based specification and post-hoc verification of imperative legacy code. They draw  their experience from two case studies, namely the PikeOS microkernel~\cite{kaiser2007evolution} verified with VCC~\cite{cohen2009vcc} and the sElect voting system~\cite{kuesters2011verifiability} verified with KeY. Baumann et al.\ \cite{baumann2012lessons} report on their experience of the verification tasks in the Verisoft XT project. They also verified the PikeOS microkernel using VCC and discuss challenges they encountered in bridging informal and formal specification. Gouw et al.\ \cite{de2015openjdk} investigated the correctness of TimSort with KeY. They indeed discovered a bug in its implementation and derived a bug-free implementation that was proven correct. Beckert et al.\ \cite{beckert2017proving} conducted another case study by formally specifying JDK's dual pivot quick sort method with JML and proving it correct in KeY. Estler et al.\ \cite{estler2014contracts} present  a study that investigates how contracts are used in the practice of software development. They analyzed a total 21 projects in the programming languages Java, Eiffel, and C\#, which all were following the design-by-contract methodlogy to some extent. Pariente and Ledinot~\cite{pariente2010formal} conducted a case study on formal verification of industrial C code using the verification system Frama-C~\cite{cuoq2012frama}. Their results are in-line with ours, as they vividly exhibit that deductive verification of industrial source code requires considerable expertise.

%% file: content/conclusion.tex
\section{Conclusion and Future Work}
\label{sec:conclusion}

Driven by research, formal verification of highly complex software systems made considerable progress in the last decades. Beyond its purpose to increase trust in the correctness of a program, it also prevents safety-related bugs, where life or missions are at stake. Furthermore, better tool support and advanced automation push contract-based verification even in the range of industrial software developers, which do not have to be highly specialized to apply verification techniques. However, although scalability of verification techniques is addressed significantly, scalability of the specification phase is seldom investigated and discussed.

In our real-world case study, we specified parts of OpenJDK's Collections-API with JML and verified them with the deductive verification system KeY. Our approach was to take the perspective of an inexperienced user to gain insights about challenges that a typical software developer would face. We described issues that occurred during the specification with JML and the verification with KeY and tried to present ideas that would resolve them. Our vision here is to aid developers of deductive verification tools to make them applicable to industrial software developers.

One disillusioning insight of this case study was that deductive verification still requires high expertise in the underlying proof theory. The reasons are manifold. First, if a proof cannot be closed, identifying whether the problem lies in the specification or implementation was notoriously hard, even for simple methods. Second, there exist a considerable amount of parameters to set for verification tools and particularly for KeY, each with the possibility to influence provability and verification effort. However, finding the right configuration without feedback and tool support is impractical. Third, based on Java's underspecification of the maximal array length, verification systems may verify a method that can also fail when exploited by malicious software. This issue raises awareness of the fact that formal verification and software testing should be applied in concert to increase trust in the correctness.

For deductive verification tools to become applicable by typical software developers, we believe that raising awareness of the challenges in the specification and verification process is necessary. Hence, a community effort is needed to specify widely-used APIs such as OpenJDK that users can verify their own software against. There are several directions to extend this work.

%Claims about increased effectiveness or productivity attributed to new methods or tools
%should be backed up by experimental user studies. This should answer such questions as
%in which manner interactive verification needs to be more automated.

\begin{itemize}
\item Besides specifying and verifying a larger part of OpenJDK to gain more experiences, it is necessary to also employ other deductive verification systems, preferably for numerous programming and specification languages. 
%Examples are . 
This allows us to generalize some of our findings and identify common challenges in terms of scalability of the specification process and usability of the verification environment. 
\item Accordingly, usability of the numerous IDEs should be investigated in experimental user studies. For instance, success of verifiying source code often depends on a user-chosen parameterization, but specific parameters are hard to understand and sometimes even negligible. Tool builders need to be aware of the user's challenges.
\item Additionally, identifying reliable and scalable means for inferring specifications for API-like code (semi)-automatically can ease the specification process. We identified that certain specification aspects have to be spelled out explicitly that could also be synthesized from the code -- either statically or dynamically. In particular, APIs evolve and new methods for avoiding unneeded re-verifications under change become crucial. 
\end{itemize}
%How can formal software verification be usefully integrated into a software development process?

%Develop techniques to infer specifications from code in a (semi-)automated manner. Many specification details that have to be spelled out explicitly, actually can be inferred from the code. There is initial work on specification generation, but it is not integrated into deductive verification frameworks

%An interesting field of research in this regard is whether contracts can be inferred from the implementation. We also believe that we need to come to a point where full-automatic verification of less complex software is generally possible. 